\documentclass{webofc}

\usepackage[varg]{txfonts}   
\usepackage{hyperref}
\usepackage{url}
\usepackage{epsfig}
\hypersetup{colorlinks=true,citecolor=blue,urlcolor=blue,linkcolor=blue}
\begin{document}
\title{Top-mass determination from leptonic final states}
%
%

\author{\firstname{Gennaro} \lastname{Corcella}\inst{1}\fnsep\thanks{\email{
  gennaro.corcella@lnf.infn.it}}}

\institute{INFN, Laboratori Nazionali di Frascati, Via E.~Fermi 54, I-00044 Frascati (RM), Italy}

\abstract{The top-quark mass is a fundamental parameter of the Standard Model,
  as it plays a crucial role in the electroweak
  precision tests, stability of the
  vacuum and inflation. I review the method and the main results contained
  in a recent ATLAS analysis which measures the top mass by using the invariant
  mass of the leptons coming from $W$ and $B$-hadron decays.
  The extracted top mass turns out to be
  the most precise single measurement by ATLAS.
}
\maketitle
\section{Introduction}

The mass of the top quark ($m_t$) is a fundamental quantity of the
Standard Model of strong and electroweak interactions,
which enters in the electroweak precision tests and constrained
the value of the Higgs boson mass even before its discovery. 
Furthermore, it is of paramount importance in inflation and in the
determination of the stability of the universe.

Since the top quark is
the only quark which decays before hadronization, the interpretation
of the top-mass measurements has been the subject of a long-standing debate,
since the most accurate measurements have been obtained by using
Monte Carlo event generators, which are not exact QCD calculationss.
Much work has been carried out in addressing
the uncertainty/discrepancy of the extracted mass when expressed in terms
of well-defined field theory definitions, such as the pole mass
(see, e.g., \cite{corcella,nason,hoang} for reviews).
It is well understood that the mass obtained from direct measurements
relying on top decays ($t\to bW$)
must be close to the pole mass, up to corrections
which depend on the top width or non-pertubative effects, such as colour
reconnection or underlying event.
Having in mind this discussion on the nature of the measured top mass,
this contribution  deals with
a recent top-mass measurement carried out at the LHC
by the ATLAS Collaboration \cite{atlas} at a centre-of-mass
energy $\sqrt{s}=13$~TeV and
integrated luminosity ${\cal L}=36$~fb$^{-1}$.
This analysis uses leptonic final states in top decays and
the so-called soft-muon tagging (SMT), i.e., it exploits
the invariant mass
$m_{\ell\mu}$, where $\ell$ is a charged
lepton from the $W$ and $\mu$ comes from the
decay of a $b$-flavoured hadron.
The advantage of this method is that, as it is based on fully-leptonic
final states, it exhibits a mild sensitivity to jet calibration and
to the $t\bar t$ production phase; on the other hand, one
should expect a major dependence
on $b$-quark fragmentation, as it relies on the description
of the hadronization of a $b$ into a $B$-hadron, eventually decaying into
a muon.
Nevertheless, as will be detailed, such an analyis yields the
most precise single $m_t$ measurement by ATLAS.

In the following section, I will give some insight on the analysis
\cite{atlas}, paying special attention to the Monte Carlo modelling and
uncertainties in the
top mass measurement. I will finally make some concluding remarks.

\section{Measuring the top mass using the soft-muon tagging}
The analysis presented hereafter makes use of data collected in 2015 and 2016 by
the ATLAS Collaboration at the LHC, namely centre-of-mass energy 
$\sqrt{s}=13$~TeV and integrated luminosity ${\cal L}=36.1$~fb$^{-1}$.
The selected $t\bar t$ events feature lepton+jets final states, i.e.
one $W$ decays leptonically ($W\to \ell\nu$) and the other one into
jets ($W\to jj'$). At parton level, one has a final state
$b\ell\nu\bar bjj'$, i.e. four jets, one charged lepton and missing
energy due to the neutrino.
Furthermore, as pointed out in the introduction,
at least one of the $b$-flavoured jets
in $t\to bW$ is required to yield 
a muon $\mu$, coming from the semileptonic decay of a $B$-hadron.
The invariant mass $m_{\ell\mu}$ is then employed to extract $m_t$.

Final-state jets are clustered according to the anti-$k_T$ algorithm 
\cite{antikt} with a radius parameter $R=0.4$ and must
have rapidity $|\eta_j|<2.5$ and transverse momentum $p_{T,j}>25$~GeV.
As for primary leptons, i.e. those coming from $W$ decays,
the corresponding cuts are $p_{T,\ell}>27$~GeV and $|\eta_\ell|<2.5$. 
One also sets an invariant opening angle between jets and primary leptons
$\Delta R_{j\ell}>0.4$, where $\Delta R=\sqrt{(\Delta \eta)^2+(\Delta\phi)^2}$,
$\phi$ being the azimuthal angle. 
In the selected sample each
event must have at least one SMT, namely a $b$-jet
where the $B$-hadron decays into a soft muon.
The cuts applied on such a muon are $p_{T,\mu}>8$~GeV, $|\eta_\mu|<2.5$
and an angular distance $\Delta R_{j\mu}<0.4$ from its own jet candidate.
Furthermore, one requires $\Delta R_{\ell\mu}<2$ between primary leptons and soft
muons, in such a way to enhance the probability that both leptons come
from the same top and minimize the fraction of events
where they instead originate from two different top quarks. 

The simulation of the $t\bar t$ samples is carried out by using
the $hvq$ setup of POWHEG \cite{powheg} with the NLO NNPDF3.0 
parton distribution functions \cite{nnpdf}. Off-shell
and width effects are included in the MADSPIN approximation,
based on Ref.~\cite{madspin}, while
showers, underlying event and
hadronization are modelled by the PYTHIA 8.2 code \cite{pythia}.
HERWIG 7.1.3 \cite{herwig} is instead used for the sake of comparison
and estimating the Monte Carlo systematics.
Bottom- and charm-hadron decays and mixing are handled by the
EVTGEN 1.2.0 program \cite{evtgen}, with the production fractions
and branching ratios rescaled to agree with the PDG values \cite{pdg}.
In the $hvq$ POWHEG framework, NLO QCD corrections are applied 
to $t\bar t$ production, while  gluon radiation in
top decays is completely handled by the PYTHIA shower. Matrix-element
corrections to parton showers in top decays,
namely NLO tree-level corrections 
$t\to bWg$, are included along the lines of \cite{mepythia}.
The $h_{\rm damp}$ parameter in POWHEG, ruling the
transverse momentum of the first emission, is set to
$h_{\rm damp}=1.5~m_t$; moreover,
the total $t\bar t$ cross section is rescaled to
the NNLO value computed by the Top++ code \cite{toppp}, i.e.
$\sigma_{t\bar t}\simeq ( 832^{+46}_{-51})$~pb.
Backgrounds to $t\bar t$ production are simulated by
SHERPA 2.2.1 \cite{sherpa} or POWHEG itself: the details
of the background simulation are discussed in \cite{atlas} and
are not reported here for the sake of brevity.

\begin{figure}[t]
\begin{center}
{\epsfig{file=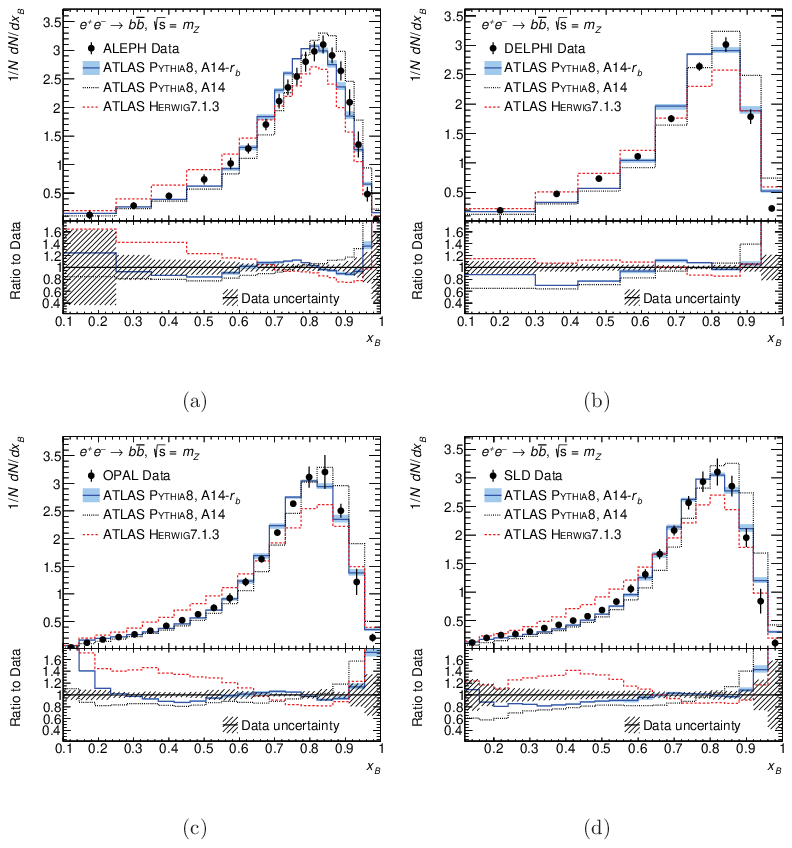,height=5.in,width=4.in}}
\end{center}
\vspace{3.5cm}\label{fig-1}
\caption{Comparison of the $x_B$ data from LEP and SLD at the $Z$ pole with 
PYTHIA and HERWIG simulations.
As for PYTHIA, both A14 and A14-$r_b$ setups are employed. Figures taken
from \cite{atlas}.}
\end{figure}

The specific PYTHIA tuning employed in this analysis deserves some
thorough discussion. The ATLAS Collaboration released the A14 tuning
\cite{a14}, based on LEP and Tevatron data, as well as on a few
ATLAS measurements
at $\sqrt{s}=7$~TeV, in order to account for effects which are relevant
at hadron colliders, such as underlying event, colour reconnection
or multiple interactions. However, since this top-mass determination
crucially depends on the fragmentation of $b$ quarks into $B$-hadrons,
the ATLAS Collaboration decided to retune the PYTHIA $b$-fragmentation
parameters, for the sake of a reliable measurement.
In PYTHIA, $b$-quark hadronization is described according to the 
so-called
Lund-Bowler non-perturbative fragmentation function \cite{lund,bowler}:
\begin{equation}\label{bowler}
  f(z)=\frac{1}{z^{1+br_bm_b^2}}(1-z)^a\exp\left(-bm_T^2/z\right).
  \end{equation}
In Eq.~(\ref{bowler}) $m_b$ is the $b$-quark mass, $m_T$ the
$B$-hadron transverse mass, i.e. $m_T=\sqrt{p_T^2+m_B^2}$,
$z$ the hadron longitudinal energy fraction with respect to the parent
quark. $a$, $b$ and $r_b$ are instead free parameters which
must be tuned to experimental
data. However, while $a$ and $b$ are sensitive to light-hadron production,
$r_b$ is specific to $b$-quark fragmentation.
Therefore, in \cite{atlas} $a$ and $b$ are kept to their default values,
while $r_b$, whose default is $r_b=0.85$ in A14, is instead retuned to data
from LEP (ALEPH \cite{aleph}, DELPHI \cite{delphi}
and OPAL \cite{opal} experiments) and SLD \cite{sld}
on $B$-hadron production in $e^+e^-$ annihilation at the $Z$ pole
\footnote{$b$-quark fragmentation in Refs.~\cite{aleph,delphi,opal,sld}
  is described in terms of the energy fraction
  $x_B=2 p_B\dot p_Z/m_Z^2$,
  $p_B$ and $p_Z$ being the four-momenta of the $B$-hadron and of the $Z$ boson, respectively.}.
The details of the fit are described in \cite{atlas} and lead to
the value $r_b=1.05\pm 0.02$: the obtained setting is referred
to as A14-$r_b$. The comparison of the tuned PYTHIA simulation with the
$e^+e^-$ data is presented in Fig.~\ref{fig-1}, which also contains the
predictions of the HERWIG 7.1.3 event generator \cite{herwig}:
the improvement of the A14-$r_b$ setup with respect to the default one is
well visible. Before presenting the results on the top-mass extraction,
it should be pointed out that, in principle, using
in $pp$ collisions a Monte Carlo fit to $e^+e^-$ data may be arguable,
since in a hadron environment one has effects like initial-state radiation,
underlying event or colour reconnection between initial and final state, which
do not have a counterpart in lepton colliders. From this point of view,
a fit to precise $pp$ data would be preferable. However, the analysis setup
of \cite{atlas} indeed minimizes such effects, as one just deals with leptonic
final states, and the standalone A14 fit includes information
coming from LHC data. Furthermore, in the narrow-width
approximation for the top quark, 
shower and hadronization of $b$ quarks in $t\to bW$ are expected to be
similar to $Z\to b\bar b$, and hence a tuning of $r_b$ to LEP and SLD is
expected to be reliable even for $b$-quark fragmentation in top decays. 
\par
\begin{figure}[t]\hspace{4.cm}\label{fig-2}
{\epsfig{file=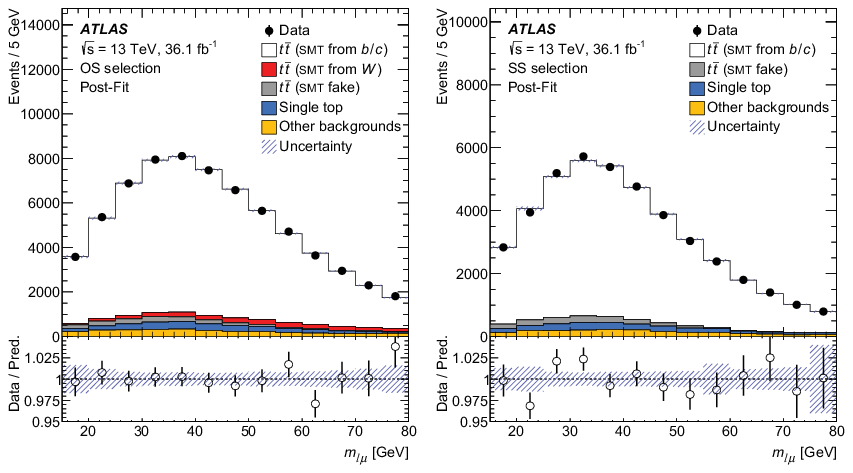,angle=90,angle=90,angle=90,height=3.in,width=5.in}}
\caption{Data on the invariant mass $m_{\ell\mu}$ compared with
  tuned POWHEG+PYTHIA for $t\bar t$ signal and background as
  discussed in \cite{atlas}. The OS and SS labels refer to whether
  the lepton from the $W$ and the soft muon from the $B$-hadron
  have opposite (OS) or same (SS) sign. The uncertainty band includes
statistical and systematic uncertainties. Figures taken from \cite{atlas}.}
\end{figure}

 The top mass is then extracted by comparing the POWHEG+PYTHIA prediction,
based on the A14-$r_b$ tuning, with the data on the $m_{\ell\mu}$ invariant
mass, as displayed in Fig.~\ref{fig-2}, for distributions obtained after the
fit.
In \cite{atlas}, checks are carried out fitting separately the electron and
muon channels, as well as the samples where the leptons from $W$ and $B$ have
the same (SS) or opposite (OS) signs.
The fit yields the following result:
\begin{equation}\label{mtfit}
  m_t=174.41\pm 0.39~({\rm stat.})\pm
0.66~({\rm syst.)} \pm 0.25~({\rm recoil}).\end{equation}
While for a discussion on the determination
of the systematic and statistical errors
one can refer to \cite{atlas}, the so-called recoil uncertainty
quoted in (\ref{mtfit}) deserves further comments.
In fact, in the treatment of the PYTHIA parton shower off the $b$ quark in
$t\to bW$, one has the option to change the gluon recoil scheme from
recoiling against the $b$ (often labelled as RTB),
which is the default setting, to recoiling against the $W$ (RTW) or the
top (RTT). The RTW and RTT options typically lead to wide-angle radiation
in top decay and deposits of energy out of the clustered $b$-jet, while
the RTB one is more consistent with the shower collinear approximation.
As a matter of fact, in \cite{atlas} the change of the recoil
modelling is assumed to be another source of uncertainty.
Some hints on the reliability of the various options can be obtained by
comparing the PYTHIA results for the $B$-hadron energy fraction in top decays,
defined in an equivalent way to $x_B$ in Fig.~1, with the NLO+NLL
resummed calculation of Ref.~\cite{cacciari}, which uses the Kartvelishvili
non-perturbative fragmentation function \cite{kart}, tuned to
ALEPH data \cite{aleph} to model hadronization. Ref.~\cite{cacciari} also
investigates bottom-quark fragmentation in top decays in Mellin moment space,
including the non-perturbative information by means of a fit to DELPHI
data  in $N$-space \cite{delphi}.
Overall, it is found that the RTB setting data gives the best comparison with
\cite{cacciari}, while the RTT and RTW options lead to a shift in the
extracted mass of about 0.25 GeV.
Therefore, Ref.~\cite{atlas} quotes a recoil uncertainty of 0.25 GeV,
as in Eq.~(\ref{mtfit}), in the measured top mass with the SMT method.

\section{Conclusions}
I reviewed a recent analysis undertaken by the ATLAS Collaboration
on the top-mass determination by using the invariant mass of a lepton
and a soft muon, coming from a $W$ and a $B$-hadron in top decays,
respectively. This measurements fully relies on 
leptonic final states in top decays, which minimizes uncertainties
due to the $t\bar t$ production stage and jet energy scale and calibration.
For the sake of a reliable measurement, the bottom-quark fragmentation
parameter $r_b$ in the PYTHIA event generator was retuned to LEP and SLD data.
As a matter of fact,
for the time being, this top-mass extraction turns out to be
the most precise single measurement from the reconstruction of the
top-decay products, i.e. $m_t=174.41\pm 0.39~({\rm stat.})\pm
0.66~({\rm syst.)} \pm 0.25~({\rm recoil})$.
While the determination of the systematic and statistical uncertainties
is rather standard, particular care was taken about the assessment
of the one due to the choice of the PYTHIA recoil scheme for
gluon radiation in top decays. For this purpose,
the comparison with a resummed calculation for $b$-quark fragmentation in top
decays turned out to be very useful to address the best option for the
recoil scheme and give a meaningful estimate of the uncertainty. 
Overall, the presented top-mass measurement is in agreement with other
determination and, thanks to the negligible error due to the jet activity,
it is expected to play a relevant role in the future combinations.

\section*{Acknowledgements}
The presented work was carried out during the association of the
author with the ATLAS Collaboration as an ACE (Analysis Consultant and Expert).
The author wishes to warmly thank Lucio Cerrito, who proposed the association
with the Rome Tor Vergata ATLAS group,
as well as the other members of the analysis 
team Veronique Boisvert, Umberto De Santis, Francesco Giuli, Michele Pinamonti 
and Marco Vanadia.

%
%

\end{document}